\documentstyle[eqsecnum,aps,twocolumn]{revtex}

\begin{document}
\author{H. K. Avetissian, G.F.Mkrtchian, M.G. Poghosyan}
\address{Department of Theoretical Physics, Plasma Physics Laboratory,Yerevan State
University\\
1, A. Manukian, 375049 Yerevan , Armenia\\
E-mail: avetissian@ysu.am\\
Fax: (3741) 570--597, Phone: (3741) 570--597}
\title{Relativistic Quantum Theory of Cyclotron Resonance in a Medium }
\maketitle

\begin{abstract}
In this paper the relativistic quantum theory of cyclotron resonance in an
arbitrary medium is presented. The quantum equation of motion for charged
particle in the field of plane electromagnetic wave and uniform magnetic
field in a medium is solved in the eikonal approximation. The probabilities
of induced multiphoton transitions between Landau levels in strong laser
field is calculated.
\end{abstract}

\section{INTRODUCTION}

As is known when charged particle moves in the field of transverse
electromagnetic wave (EMW) in the presence of uniform magnetic field
directed along the wave propagation direction, a resonant effect of the wave
on the particle motion is possible. If the interaction takes place in vacuum
this is the well known phenomenon of autoresonance \cite{K1}-\cite{K2} when
the ratio of the Doppler shifted wave frequency ($\omega ^{^{\prime }}$) to
the cyclotron frequency of a particle ($\Omega $) is conserved $\omega
^{^{\prime }}/\Omega =const$ and the resonance created at the initial moment
is automatically supported throughout interaction. However, if the
interaction takes place in a medium where the phase velocity of a EMW larger
(plasma like medium) or smaller (dielectric medium) than light speed ($c$)
the picture of wave-particle interaction is essentially changed.
Particularly, in a medium the autoresonance phenomenon is violated because
of nonequidistant Stark-shift of magnetic sublevels of an electron (Landau
levels) in the electric field of EMW. As a result the intensity effect of
the wave governs the resonance characteristics, and the particle state
essentially depends on the initial conditions and the wave field magnitude,
at which the nonlinear resonance is achieved \cite{Av1}. At first the
investigation of cyclotron resonance (CR) in the medium in the scope of
classical theory was carried out in the papers \cite{K2},\cite{Buch}, where
oscillating solutions for the particle energy are obtained. However, such
behavior is valid only for EMW intensity less than some critical value. As
was shown in \cite{Av1}, for larger intensities a non-linear resonance
phenomenon of so called ''Electron Hysteresis'' takes place when EMW is
turned on adiabatically. If the intensity pick of an actual wave pulse
exceeds the mentioned critical value then significant acceleration of
charged particles can be achieved (medium may be either plasma or refractive)%
\cite{Av1}, \cite{Avc}.

Concerning the quantum description of CR the relativistic quantum equation
of motion allows exact solution only for CR in vacuum \cite{Red}
(description of related Quantum Electrodynamic (QED) processes, such as
electron-positron pair production, non-linear Compton scattering in the
presence of uniform magnetic field and etc. by this wave function see \cite
{Terb} and references therein). It is worthy to mention that the
configuration of electromagnetic fields when uniform magnetic field is
directed along the propagation of transverse wave is one of the exotic cases
when the relativistic quantum equation of motion allows exact solution. In
the medium even at the absence of uniform magnetic field the relativistic
quantum equation of motion describing the particle-wave interaction reduces
to the Mathieu type (in general, Hill type) equation the exact solution of
which is unknown. In this case to obtain an approximate analytical solution
describing the nonlinear process of interaction is already problematic \cite
{Av2},\cite{Av3}.

The purpose of this paper is to obtain non-linear over the EMW field
approximate solution of relativistic quantum equation of motion for a
charged particle in the strong EMW in the medium, in the presence of uniform
magnetic field, which will good enough describe the quantum picture of
cyclotron resonance, particularly multiphoton stimulated transitions between
Landau levels. By this wave function one can treat a large class of
non-linear QED processes in strong electromagnetic fields with modifications
that the medium brings (e. g., anomalous Doppler effect), taking into
account Astrophysical applications as well where CR plays a significant role 
\cite{Ginzb}. One of the advantages of CR in the refractive medium is that
for moderate relativistic particle beam one can achieve the cyclotron
resonance in optical region (close to Cherenkov resonance) by current lasers
and existing uniform magnetic fields (\symbol{126}$10^4G$), while in the
vacuum for the same parameters CR is possible for radio frequencies. The
Free Electron Laser version based on the combine scheme of CR and Cherenkov
radiation was purposed in \cite{Av4}.

This paper is organized as follows. In Sec. II the wave function of a
charged particle moving in a medium in the field of transverse EMW, at the
presence of uniform magnetic field directed along the wave propagation
direction, is obtained. In Sec.\ III the CR is considered in the medium and
the probabilities of induced multiphoton transitions in the strong
circularly polarized EMW are calculated.

\section{Wave Function of a Particle in a Plane EMW in the Medium at the
Presence of Uniform Magnetic Field}

Let charged particle moves in a medium in the following configuration of EM
field

\begin{equation}
A=A_H+A_w  \label{A}
\end{equation}
where 
\begin{equation}
A_H=(0,xH_0,0,0),  \label{Ah}
\end{equation}
is the four-vector potential of uniform magnetic field with the strength $%
H_0 $ directed along the $z$ axis and 
\begin{equation}
A_w=\left\{ A_x\left( t-n\frac zc\right) ,A_y\left( t-n\frac zc\right)
,0,0\right\}  \label{Aw}
\end{equation}
is the four-vector potential of transverse EMW propagating along the $z$
axis. Here for the four-component vectors we have chosen the following
metric $a=(\overrightarrow{a},ia_0)$. In (\ref{Aw}) $n$ is the refraction
index of the medium (it is assumed quasimonochromatic wave: $n(\omega
)\simeq n$) and $c$ is the light speed in vacuum. We will assume that the
EMW is switched on/off adiabatically so for the four-vector potential we
have $A_w=0$ at $t=\pm \infty $ .

In present consideration we will restrict the total energy exchange $\Delta
E $ of a particle with EMW

\begin{equation}
\Delta E<<E,  \label{2a}
\end{equation}
where $E$ is the particle energy.

From the condition (\ref{2a}) follows the restriction on the wave frequency $%
\omega $:

\begin{equation}
\hbar \omega <<E\text{ }.  \label{2}
\end{equation}
For this reason in considering case the neglecting of the spin interaction
is justified. So we will use the Klein-Gordon equation which for a charged
particle in the field (\ref{A}) is

\begin{equation}
\left\{ \left( i\hbar \partial _\mu +\frac ecA_\mu \right) ^2+m^2c^2\right\}
\Psi =0,  \label{3}
\end{equation}
where $\hbar $ is the Plank constant $m$ and $e$ are the particle mass and
charge respectively and $\partial _\mu \equiv \partial /\partial x_\mu $ $%
(\mu =1,2,3,4)$ denotes the first derivative of a function with respect to
four-component radius vector $x$.

The charged particle initial state at $t$ $\rightarrow -\infty $ when $A_w=0$
is well known and has been the topic of numerous studies (see, e.g.,\cite
{Akh}). As it is known \cite{Akh} the motion of the particle in the uniform
magnetic field is separated into cyclotron ($x,y$) and the longitudinal ($z$%
) degrees of freedom. For longitudinal motion we will assume initial state
with momentum $p_z$, while for cyclotron motion we will assume the state \{$%
s,p_y$\}, where by $s$ we indicate Landau levels and by $p_y$ the $Y$
component of generalized momentum. So the particle initial state when the
EMW is adiabatically switched on at $t$ $\rightarrow -\infty $ is assumed to
be

\begin{equation}
\psi _s=N\Phi _s(x)\exp \left[ \frac i\hbar (p_zz+p_yy-E_s(p_z)t)\right] ,
\label{1}
\end{equation}
where $N$ is the normalization constant,

\begin{equation}
\Phi _s(x)=\frac 1{\sqrt{2^ss!a\sqrt{\pi }}}\exp \left[ -\frac{\left( x-%
\frac{cp_y}{eH_0}\right) ^2}{2a^2}\right] U_s\left[ \frac{x-\frac{cp_y}{eH_0}%
}a\right]  \label{ermit}
\end{equation}
are the wave functions corresponing to Landau levels. Here $U_s$ are Hermit
polynomials with

\[
\quad a=\sqrt{\frac{e\hbar H_0}c} 
\]
and dispersion law is 
\begin{equation}
E_s^2(p_z)=m^2c^4+c^2p_z^2+2ecH_0\hbar \left( s+\frac 12\right) \text{.}
\label{en}
\end{equation}

As the EMW field depends only on the $\tau =t-\frac ncz$ then raising from
the symmetry, the particle wave function can be looked for in the following
form:

\begin{equation}
\Psi ({\bf r},{\bf t})=f(x_{\bot },\tau )\exp \left[ \frac i\hbar
(p_zz-Et)\right]  \label{4}
\end{equation}
Taking into account (\ref{2a}) we can consider $f(x_{\bot },\tau )$ as a
slowly varying function of $\tau $ and neglect the second derivative
compared with the first order. So from (\ref{3}) for $f(x_{\bot },\tau )$ we
will have the following equation:

\begin{equation}
\left\{ \frac{2i\hbar }{c^2}\widetilde{E}\partial _\tau -\left( i\hbar
\partial _\mu ^{\bot }+\frac ecA_\mu \right) ^2+\frac{E^2}{c^2}%
-m^2c^2-p_z^2\right\} f=0,  \label{6}
\end{equation}
where 
\[
\partial _\mu ^{\bot }=\left\{ \partial _x,\partial _y,0,0\right\} ,\qquad
x_{\bot }=\left\{ x,y,0,0\right\} ,\qquad \widetilde{E}=E-cnp_z 
\]
In Eq. (\ref{6}) transverse and longitudinal motions are not separated. But
after a certain unitarian transformation in the equation for the transformed
function the variables are separated \cite{Terb}, that is 
\begin{equation}
\widehat{S}=\exp \left\{ i{\bf K}(\tau )\widehat{{\bf P}}_{\bot }\right\}
;\qquad \widehat{P}_{\bot \mu }=-i\hbar \partial _\mu ^{\bot }-\frac ec%
A_{H\mu }\text{ ,}  \label{7}
\end{equation}
where ${\bf K}(\tau )$ will be chosen to separate the cyclotron and
longitudinal motions and to fulfil the initial condition (\ref{1}):

\[
K_x+iK_y=-\exp \left[ -i\frac{ec}{\widetilde{E}}H_0\tau \right] 
\]

\begin{equation}
\times \int_{-\infty }^\tau \frac{ec}{\hbar \widetilde{E}}\left( A_x\left(
\tau ^{\prime }\right) +iA_y\left( \tau ^{\prime }\right) \right) \exp
\left[ i\frac{ec}{\widetilde{E}}H_0\tau ^{\prime }\right] d\tau ^{\prime }
\label{8}
\end{equation}
For the transformed wave function $\widetilde{f}=\widehat{S}f(x_{\bot },\tau
)$ we will have the following equation 
\[
\left\{ -\frac{E^2}{c^2}+p_z^2+m^2c^2-i\frac{2\hbar \widetilde{E}}{c^2}%
\partial _\tau -\frac{e\hbar ^2\widetilde{E}}{c^3}K^\nu F_{\nu \mu }\frac{%
dK^\mu }{d\tau }\right. 
\]

\begin{equation}
\left. +\widehat{{\bf P}}_{\bot }^2+\left( \frac{e\hbar }cF_{\mu \upsilon
}K^\upsilon {\bf e}^\mu +\frac ec{\bf A}_w\right) ^2\right\} \widetilde{f}%
(x_{\bot },\tau )=0,  \label{9}
\end{equation}

Where $F_{\mu \upsilon }$ is the tensor of EM field corresponding to uniform
magnetic field and ${\bf e}^\mu =\left\{ 1,1,0,0\right\} $. In Eq. (\ref{9})
the variables are separated and making inverse transformation $f=\widehat{S}%
^{+}\widetilde{f}(x_{\bot },\tau )$ gives the solution of the initial
equation (taking into account Eq. (\ref{4})):

\[
\Psi ({\bf r},{\bf t})=N\exp \left[ \frac i\hbar (p_zz-E_s(p_z)t)-\frac i%
\hbar \int_{-\infty }^\tau Q(\tau ^{\prime })d\tau ^{\prime }\right] 
\]
\begin{equation}
\times \exp \left[ i\frac ecH_0K^y(x-\frac \hbar 2K^x)\right] T_s\left( {\bf %
x}_{\perp }-\hbar {\bf K}\right)  \label{10}
\end{equation}
where 
\begin{equation}
T_s\left( x_{\perp }\right) =\exp \left\{ i\frac{p_yy}\hbar \right\} \Phi
_s(x)  \label{11}
\end{equation}
and 
\begin{equation}
Q(\tau )=\frac{c^2}{2\widetilde{E}}\left[ \left( \frac{e\hbar }cF_{\mu
\upsilon }K^\upsilon {\bf e}^\mu +\frac ec{\bf A}_w\right) ^2-\frac{e%
\widetilde{E}\hbar ^2}{c^3}K^\nu F_{\mu \nu }\frac{dK^\mu }{d\tau }\right]
\label{12}
\end{equation}

\section{The Probabilities of Multiphoton Transitions}

Although the motion of the particle in the uniform magnetic field is
separated into cyclotron ($x,y$) and the longitudinal ($z$) degrees of
freedom (\ref{1}), in energy scale these motions are not separated due to
relativistic effects(\ref{en}). However, for not so strong magnetic fields
we can separate the energies of longidudinal ($E_{\Vert })$ and cyclotron
motions

\begin{equation}
E_s(p_z)\simeq E_{\Vert }+\hbar \Omega \left( s+\frac 12\right) ;\quad \hbar
\Omega s<<E_{\Vert }  \label{13}
\end{equation}
\[
\Omega =ecH_0/E_{\Vert },\quad E_{\Vert }=\sqrt{m^2c^4+c^2p_z^2} 
\]

Now let us consider the concrete case of circularly polarized EMW

\begin{equation}
A_w\left( \tau \right) =\left\{ -A\left( \tau \right) \sin (\omega \tau
),gA\left( \tau \right) \cos (\omega \tau ),0,0\right\}  \label{15}
\end{equation}
wich is in resonance with the particle, i.e. Doppler shifted wave frequency
is close to cyclotron one 
\begin{equation}
\omega ^{\prime }\equiv \left( 1-nv_z/c\right) \omega \simeq g\Omega
\label{14}
\end{equation}
where $v_z$ is the particle longitudinal velocity. In (\ref{15}) $g=\pm 1$
correspond to right and left hand circular polarizations of the wave. After
the interaction ($\tau \rightarrow +\infty $ ) from (\ref{8}) at the
resonance condition (\ref{14}) we have 
\begin{equation}
K_x=-\frac{e\overline{A}cT}{\hbar E_{\Vert }}\cos (\omega \tau )  \label{16}
\end{equation}
\begin{equation}
K_y=g\frac{e\overline{A}cT}{\hbar E_{\Vert }}\sin (\omega \tau )  \label{17}
\end{equation}
where $\overline{A}$ is the average value of $A\left( \tau \right) $ and $T$
is the coherent interaction time.

The final state of the particle after the interaction is described by the
wave function 
\[
\Psi _s=N\exp \left[ \frac i\hbar (p_zz+p_yy-E_s(p_z)t)+i\frac{eg\overline{A}%
\Omega Tx}{\hbar c}\sin (\omega \tau )\right] \Phi _s\left[ \rho \right] 
\]
\begin{equation}
\times \exp \left[ -i\frac{egH_0\hbar }c\left( \frac{e\overline{A}cT}{2\hbar
E_{\Vert }}\right) ^2\sin (2\omega \tau )+i\frac{gp_ye\overline{A}cT}{\hbar
E_{\Vert }}\sin (\omega \tau )\right]  \label{21}
\end{equation}
where 
\begin{equation}
\rho =\frac 1a\left( x+\frac{e\overline{A}cT}{E_{\Vert }}\cos (\omega \tau )-%
\frac{a^2p_y}\hbar \right)  \label{22}
\end{equation}
Expanding the wave function (\ref{21}) in terms of the full basis of the
particle eigenstates (\ref{1}) 
\begin{equation}
\Psi _s=\int dp_y^{\prime }dp_z^{\prime }\sum_{s^{\prime }}C_{ss^{\prime
}}(p_y^{\prime },p_z^{\prime })\psi _{s^{\prime },p_y^{\prime },p_z^{\prime
}}  \label{24}
\end{equation}
we will find the probabilities of the multiphoton induced transitions
between the Landau levels.

To calculate the expansion coefficients $C_{ss^{\prime }}(p_y^{\prime
},p_z^{\prime })$ we will take into account the result of the following
integration 
\[
\int \exp (-ikx)\Phi _s(a^{-1}x+ab)\Phi _{s^{\prime }}(a^{-1}x+ab^{\prime }) 
\]
\begin{equation}
=\exp \left\{ i\mu +i(s-s^{\prime })\lambda \right\} I_{ss^{\prime }}(\zeta )
\label{28}
\end{equation}
where $I_{ss^{\prime }}(\zeta )$ is the Lagger functions and characteristic
parameters are determined by the expressions 
\[
\mu =\frac{ka^2(b+b^{\prime })}2;\quad \lambda =\tan ^{-1}\frac k{b^{\prime
}-b} 
\]
\[
\zeta =a^2\frac{k^2+(b-b^{\prime })^2}2\text{ .} 
\]

Then we get the following transition amplitudes, 
\[
C_{ss^{\prime }}(p_y^{\prime },p_z^{\prime })=\delta (p_y-p_y^{\prime
})\delta (p_z-p_z^{\prime }-(s-s^{\prime })g\omega n\hbar c^{-1}) 
\]
\begin{equation}
\times \exp \left\{ -\frac i\hbar (E-E^{\prime }-(s-s^{\prime })g\omega
\hbar )t\right\} I_{ss^{\prime }}\left[ \zeta \right]  \label{33}
\end{equation}
where $\delta (p)$'s are the Dirac $\delta $-functions expressing the
momentum conservation laws and the argument of the Lagger functions is

\begin{equation}
\zeta =\frac{e^2\overline{A}^2T^2\Omega }{2\hbar E_{\Vert }}\text{.}
\label{ar}
\end{equation}
According to the expression (\ref{33}), transition of the particle from an
initial state \{$s,p_y,p_z$\} to a final state \{$s^{\prime },p_y^{\prime
},p_z^{\prime }$\} is accompanied by emission or absorption of $s-s^{\prime
} $ photons. Consequently, substituting (\ref{33}) into (\ref{24}) and
integrating by momentum we can write the particle wave function in another
form: 
\[
\Psi _s=N\sum_{s^{\prime }}I_{ss^{\prime }}(\zeta )\exp \left\{ -\frac i\hbar
(E-(s-s^{\prime })g\omega \hbar )t\right. 
\]
\begin{equation}
\left. +\frac i\hbar (p_z-(s-s^{\prime })g\omega n\hbar c^{-1})z+\frac i\hbar
p_yy\right\} \Phi _{s^{\prime }}\left( x\right)  \label{35}
\end{equation}
The probability of the induced transition $s\rightarrow s^{\prime }$ between
the Landau levels ultimately is defined from the formula (\ref{35}):

\begin{equation}
w_{ss^{\prime }}=I_{ss^{\prime }}^2\left[ \frac{e^2\overline{A}^2T^2\Omega }{%
2\hbar E_{\Vert }}\right]  \label{36}
\end{equation}
As is seen from (\ref{35}), when $1-nv_z/c>0$ and $g=1$ in the field of
strong EMW the Landau levels are exited at the absorption of the wave quanta
- normal Doppler effect, while in the case $1-nv_z/c<0$ and $g=-1,$ which is
possible in the refractive medium ($n>1$), Landau levels are exited at the
emission of the wave quanta, i.e. takes place anomalous Doppler effect.

Let us estimate the average number of emitted (absorbed) photons in
quasiclassical limit ($s>>1$) when multiphoton processes dominate and the
process has a classical nature. The argument of the Lagger function can be
represented as

\begin{equation}
\zeta =\frac 1{4s}\left( \frac{\triangle \varepsilon _{cl}}{\hbar \omega }%
\right) ^{2~},  \label{37}
\end{equation}
where $\triangle \varepsilon _{cl}\simeq e{\rm Ev}_{tr}T$ (${\rm E}$ is the
EMW field strength, ${\rm v}_{tr}\simeq c\sqrt{2\hbar s\Omega /E_{0\Vert }}$
is the particle mean transverse velocity) is the energy change of the
particle according to classical theory. At high intensities of the EMW: $%
\triangle \varepsilon _{cl}>>\hbar \omega $, the Lagger function is maximal
at $\zeta \rightarrow \zeta _0=\left( \sqrt{s^{\prime }}-\sqrt{s}\right) ^2$%
, exponentially falling beyond $\zeta _0$. For the transition $s\rightarrow
s^{\prime }$ and when $\left| s-s^{\prime }\right| <<s$ we have $\zeta
_0\simeq $ $\left( s^{\prime }-s\right) ^2/4s$. Comparison of this
expression with (\ref{36}) shows that the most probable transitions are

\[
\left| s-s^{\prime }\right| \simeq \frac{\triangle \varepsilon _{cl}}{\hbar
\omega } 
\]
in accordance with the correspondence principle.

\acknowledgements
This work is supported by International Science and Technology Center (ISTC)
Project No. A-353.

\end{document}